\pdfoutput=1

\documentclass[11pt,twoside,a4paper,cmspaper,final,collab]{cms-tdr}
\def\svnVersion{310274}\def\cmsCernNo{2013-037}\def\cmsCernDate{\today}\def\cmsMessage{Published in the European Physical Journal C as \href{http://dx.doi.org/10.1140/epjc/s10052-015-3706-0}{\doi{10.1140/epjc/s10052-015-3706-0.}}}

\begin{document}\cmsNoteHeader{SMP-12-016}

\hyphenation{had-ron-i-za-tion}
\hyphenation{cal-or-i-me-ter}
\hyphenation{de-vices}
\RCS$Revision: 310215 $
\RCS$HeadURL: svn+ssh://svn.cern.ch/reps/tdr2/papers/SMP-12-016/trunk/SMP-12-016.tex $
\RCS$Id: SMP-12-016.tex 310215 2015-11-12 00:52:24Z trocino $
\newlength\cmsFigWidth
\ifthenelse{\boolean{cms@external}}{\setlength\cmsFigWidth{0.85\columnwidth}}{\setlength\cmsFigWidth{0.4\textwidth}}
\ifthenelse{\boolean{cms@external}}{\providecommand{\cmsLeft}{top}}{\providecommand{\cmsLeft}{left}}
\ifthenelse{\boolean{cms@external}}{\providecommand{\cmsRight}{bottom}}{\providecommand{\cmsRight}{right}}
\ifthenelse{\boolean{cms@external}}{}{\setcounter{secnumdepth}{2}}
\ifthenelse{\boolean{cms@external}}{\providecommand{\cmsTable}[1]{#1}}{\providecommand{\cmsTable}[1]{\resizebox{\textwidth}{!}{#1}}}
\cmsNoteHeader{SMP-12-016}
\title{Measurements of the \cPZ\cPZ\ production cross sections in the $2\Pl2\PGn$ channel in proton-proton collisions at $\sqrt{s} = 7$ and $8\TeV$ and combined constraints on triple gauge couplings}
\titlerunning{\cPZ\cPZ\ production cross sections at 7 and 8\TeV and constraints on TGCs}

\author*[cern]{CMS Collaboration}

\date{\today}

\abstract{
Measurements of the \cPZ\cPZ\ production cross sections in
proton-proton collisions
at center-of-mass energies of 7 and 8\TeV
are presented.
Candidate events for the leptonic decay mode  $\cPZ\cPZ \to 2\Pl2\PGn$,
where \Pl denotes an electron or a muon,
are reconstructed and
selected from data corresponding to an integrated luminosity of
5.1\,(19.6)\fbinv at 7\,(8)\TeV
collected with the CMS experiment.
The measured cross sections,
$\sigma(\Pp\Pp \to \cPZ\cPZ) = 5.1_{-1.4}^{+1.5}\stat\,_{-1.1}^{+1.4}\syst\pm 0.1\lum\unit{pb}$ at 7\TeV, and
$7.2_{-0.8}^{+0.8}\stat\,_{-1.5}^{+1.9}\syst\pm 0.2\lum\unit{pb}$ at 8\TeV,
are in good agreement with the standard model
predictions with next-to-leading-order accuracy.
The selected data are analyzed to search for anomalous
triple gauge couplings involving the \cPZ\cPZ\ final state.
In the absence of any deviation from the standard model predictions,
limits are set on the relevant parameters. These limits are then
combined with the previously published CMS results for \cPZ\cPZ\ in
4\Pl\ final states, yielding the most stringent
constraints on the anomalous couplings.
}

\hypersetup{%
pdfauthor={CMS Collaboration},%
pdftitle={Measurements of the ZZ production cross sections in the 2 l 2 nu channel in proton-proton collisions at sqrt(s) = 7 and 8 TeV and combined constraints on triple gauge couplings},%
pdfsubject={CMS},%
pdfkeywords={CMS, physics, Standard Model, ZZ, cross section, anomalous couplings}}

\maketitle

\section{Introduction}
\label{sec:introduction}

The production of pairs of \cPZ\ bosons in proton-proton collisions is
a rare diboson process in the Standard Model (SM).
The measurement of the cross section and properties of this
process probe the self-interaction of electroweak gauge bosons.
The \cPZ\cPZ\ final state is also an
important background in searches for other interesting processes beyond the SM,
such as the production of high-mass Higgs bosons and their subsequent decay to
pairs of bosons~\cite{Pedro:phot} or supersymmetry~\cite{Hobbs:2010yg}.
Because of the non-Abelian structure of the electroweak gauge theory,
vector bosons can interact among themselves
and can couple in triplets (\eg \PW\PW\cPZ) or quartets
(\eg \PW\PW\cPZ\cPZ).
All couplings involving only bosons without electric charge
are expected to be null at tree level, leading to the absence of
triple gauge couplings for \cPZ\cPgg\cPgg, \cPZ\cPZ\cPgg,
and \cPZ\cPZ\cPZ.
An enhancement in the measured rate of \cPZ\cPZ\ production compared
to the expectation from the SM could indicate the existence of
anomalous boson couplings.

This paper presents measurements of the \cPZ\cPZ\ production cross
sections in proton-proton collisions at the LHC
at two different center-of-mass energies, 7 and 8\TeV, in the decay channel with
two charged leptons, electrons (\Pe\Pe) or muons
(\Pgm\Pgm), and a neutrino-antineutrino pair of any flavor
($\PGn\bar{\PGn}$).
The data were collected with the CMS detector at 7\,(8)\TeV,
corresponding to 5.1\,(19.6)\fbinv of integrated luminosity.

At tree level, \cPZ\cPZ\ pairs are primarily produced in the SM via
the $t$- and $u$-channels, following the annihilation of a
quark-antiquark pair in proton-proton collisions.
Because of the high gluon-gluon parton luminosity,
the $\cPg\cPg\to \cPZ\cPZ$ contribution has to be included. 
The production cross section calculated up to next-to-leading-order
(NLO) accuracy in strong coupling constant ($\alpha_\mathrm{S}$) is
expected to be $6.46^{+0.30}_{-0.21}$
($7.92^{+0.37}_{-0.24}$)\unit{pb} at 7\,(8)\TeV~\cite{bib:MCFM}, 
where the uncertainties refer only to the missing higher orders in the 
computation. These cross sections include a leading-order (LO)
computation of the $\cPg\cPg\to \cPZ\cPZ$ contribution, which is formally a
next-to-next-to-leading-order (NNLO) correction. Recently, complete
NNLO cross sections for \cPZ\cPZ\ production accompanied by jets have also 
been computed~\cite{bib:NNLOZZ}, leading to a further small increase
in cross section compared to Ref.~\cite{bib:MCFM}. However,
higher-order QCD corrections have been shown to be reduced
significantly when vetoing events where the diboson system is produced
in association with jets~\cite{Dixon:1999di, Binoth:2009wk}, 
as done in the present analysis. The NNLO QCD corrections apart from
the LO $\cPg\cPg\to \cPZ\cPZ$ contribution are thus neglected in 
our simulations and in the reference cross sections to which our
measurements are compared. Complete one-loop electroweak (EW)
corrections to massive 
vector boson pair production~\cite{Bierweiler:2013dja, Baglio:2013toa}
have also been published. The consequences of the EW corrections
for \cPZ\cPZ\ production are that the transverse momentum (\pt)
spectrum of the \cPZ\ bosons falls more rapidly and, in addition, the
overall cross section decreases by about 4\% at LHC center-of-mass
energies. 

The production of \cPZ\cPZ\ pairs has been studied at the LHC by the
ATLAS experiment, which analyzed the decay modes $2\Pl2\Pl^{\prime}$
and $2\Pl2\PGn$ ($\Pl,\Pl^{\prime}=\Pe,\Pgm$) at
7\TeV~\cite{:2012kg}, and by the CMS experiment, which
considered $2\Pl2\Pl^{\prime}$ final states ($\Pl=\Pe,\Pgm$ and
$\Pl^{\prime}=\Pe,\Pgm,\Pgt$) at 7\TeV~\cite{:2012rg} and
8\TeV~\cite{Khachatryan:2014dia}. Both experiments measured \cPZ\cPZ\
production cross sections in good agreement with the SM predictions
and set limits on anomalous triple gauge couplings (ATGCs).

The branching fraction for the $2\Pl2\PGn$ decay mode (where \Pl\
denotes only \Pe\ and \Pgm) is approximately six
times larger than that of the four-charged-lepton final state,
and the signal purity is enhanced at large values of the boson \pt,
where there is the greatest sensitivity to ATGC effects.
For this reason, the $2\Pl2\PGn$ channel has a
sensitivity comparable to that of the 4\Pl\ channel to ATGC.
The characteristic signature is an overall imbalance in the transverse
momentum of the event between the initial and the final states, which
consequently appears as missing transverse energy (\MET) in the
final state.
Although the branching fraction is large, this channel is rather
challenging due to the large contamination from background processes,
in particular the Drell--Yan (DY) process, which has a cross section
nearly five orders of magnitude larger than the signal.
If the \cPZ\ boson or the hadrons recoiling against it are not reconstructed
correctly, then an apparent \MET results and these events can resemble
the signal.
Other important sources of background are diboson processes,
\PW\PW\ and \PW\cPZ, with fully leptonic decays, and
$\PQt\bar{\PQt}$ production.

This paper presents a measurement of the \cPZ\cPZ\ production cross
section in the $2\Pl2\PGn$ channel as a function of the transverse
momentum (\pt) of the charged lepton pair. The distribution of the
dilepton \pt is sensitive to the presence of ATGCs.
Limits are computed and finally combined with
existing results from CMS in the four-charged-lepton final state.

\section{CMS detector}
\label{sec:CMS}

{\tolerance=600
The central feature of the CMS apparatus is a superconducting solenoid
of 6\unit{m} internal diameter, providing a magnetic field of
3.8\unit{T}. Within the superconducting solenoid volume are a silicon
pixel and strip tracker, a lead tungstate crystal electromagnetic
calorimeter (ECAL), and a brass and scintillator hadron calorimeter
(HCAL), each composed of a barrel and two endcap sections.
The silicon tracking system is used to measure the
momentum of charged particles and covers the pseudorapidity range
$\abs{\eta} < 2.5$, where
$\eta = -\ln{\left(\tan{\left(\theta/2\right)}\right)}$,
and $\theta$ is the polar angle of the trajectory of the particle with
respect to the counterclockwise-beam direction.
The ECAL and HCAL extend to a pseudorapidity range of $\abs{\eta} < 3.0$.
A steel/quartz-fiber Cherenkov forward detector extends the
calorimetric coverage to $\abs{\eta} < 5.0$.
Muons are measured in gas-ionization detectors embedded in the steel
flux-return yoke outside the solenoid.
\par}

The \MET is defined as the magnitude of the missing transverse
momentum or momentum imbalance, $\mathbf{\pt^{\text{miss}}}$, which
is the negative vector sum of the momenta in the plane transverse to
the beam of all reconstructed
particles (photons, electrons, muons, charged and neutral hadrons) in
the event.

A more detailed description of the CMS detector, together with a
definition of the coordinate system used and the relevant kinematic
variables, can be found in Ref.~\cite{bib:cms}.

\section{Simulation}
Several Monte Carlo (MC) event generators are used to simulate the signal
and background processes. The $\cPZ\cPZ \to 2\Pl2\PGn$ signal and
the $\PW\PW \to 2\Pl2\PGn$ and $\PW\cPZ \to 3\Pl\PGn$
background processes are simulated using \MADGRAPH 5~\cite{MADGRAPH},
as well as \cPZ+jets,
\PW+jets, and $\PQt\bar{\PQt}$+jets processes.
Single top-quark processes are simulated with
\POWHEG~\cite{Alioli:2010xd}.
In the simulation, vector bosons are allowed to decay to leptons of any
flavor (\Pe, \Pgm, \Pgt), since \Pgt\ leptons can contribute
to dielectron and dimuon final states through $\Pgt\to \Pe$ and
$\Pgt\to \Pgm$ decays.
For all these processes, the parton
showering is simulated with \PYTHIA 6~\cite{Sjostrand:2006za}
with the Z2\,(Z2*) tune for 7\,(8)\TeV
simulations~\cite{bib:underl}.

The cross section of the \cPZ\cPZ\ signal is computed with the NLO
generator \MCFM~\cite{bib:MCFM}, which includes
contributions from gluon-gluon initial states.
Since the present
cross section measurement and ATGC analysis rely on the \pt
distribution of \cPZ\ bosons, a precise prediction of this
distribution is required.
The charged dilepton \pt spectrum of $\cPZ\cPZ \to 2\Pl2\PGn$,
generated with \MADGRAPH
and interfaced with \PYTHIA for parton showering,
is found to be in good agreement with the corresponding spectrum computed at NLO in QCD with
\MCFM and therefore no differential correction for NLO QCD effects is
applied to the \MADGRAPH simulated sample.
In addition, the effect of NLO EW
corrections~\cite{Bierweiler:2013dja,Baglio:2013toa}
is taken into account
by reweighting the \cPZ\cPZ\ and the \PW\cPZ\ events as a function of the
partonic kinematic variables, and applying weights derived from the
calculations described in Ref.~\cite{Bierweiler:2013dja}.
These corrections yield an overall reduction of $4.1\%$ of the \cPZ\cPZ\ cross
section, as well as a softening of the boson \pt spectra that results
in a reduction of the differential cross section of about 20\% at \cPZ\
\pt of 300\GeV.

Simulated samples of the $\cPZ\cPZ \to 2\Pl2\PGn$ process that include
contributions from ATGCs (see Section~\ref{sec:atgc}) are produced
using the LO generator
\SHERPA~\cite{Gleisberg:2008ta}.
These samples are based on a LO matrix-element simulation
including up to two additional jets, matched to parton showers.

The parton distribution functions (PDF) are modeled with the CTEQ6L~\cite{bib:cteq} parametrization
in samples generated with \MADGRAPH, and the CT10 parametrization~\cite{bib:ct10}
in samples generated with \POWHEG and \SHERPA.
The detector response to the simulated events is modeled with \GEANTfour~\cite{bib:cms,Agostinelli:2002hh}.

\section{Event selection}
\label{sec:selection}

The signal consists of two \cPZ\ bosons, one decaying
into a pair of oppositely charged leptons
and the other to two neutrinos
that escape direct
detection. The final state is thus characterized by: a pair of
oppositely charged, isolated electrons or muons,
with an invariant mass within a \cPZ-boson mass window, no additional
leptons, and large \MET.

Events are selected using triggers that require the
presence of two electrons or two muons, with minimum \pt thresholds on
each lepton that depend on the dataset.
The trigger thresholds in the 8\TeV dataset are 17 and 8\GeV for
the leptons with higher and lower \pt, respectively. The thresholds for
the 7\TeV data samples are the same or lower. The 8\TeV data sample
also includes events that satisfy a single isolated muon trigger to
ensure the highest efficiency.
For events with two identified and isolated leptons having invariant
mass between 83.5 and 98.5\GeV and dilepton $\pt>45\GeV$, the trigger
efficiency is higher than 98\% in the dielectron channel and varies
from 94 to 98\% in the dimuon channel.
In addition, single-photon triggers or electron-muon triggers are used
to select control samples for the background determinations.

Electrons are selected inside the fiducial region of ECAL.
The electron candidates must have a minimum \pt of 20\GeV, and
satisfy standard identification criteria,
based on shower shape, track quality, cluster track matching, in
order to reject misidentified hadrons~\cite{Khachatryan:2015hwa}.

The muons
are selected inside the
fiducial region of the muon spectrometer,
with a minimum \pt of 20\GeV, and satisfy standard identification criteria
based on track information and isolation~\cite{Chatrchyan:2012xi}.

Events are selected
if they include a pair of same-flavor, oppositely charged leptons
that pass the identification and isolation criteria.
In order to suppress backgrounds that do not include a \cPZ\ boson,
the lepton pair is required to have an invariant mass
compatible with the \cPZ-boson mass,
between 83.5 and 98.5\GeV.
The \pt of the dilepton pair
is required to be
greater than 45\GeV. This requirement is particularly effective at
reducing the DY background because the \cPZ\ bosons produced in
\cPZ\cPZ\ events have, on average, larger \pt
than those from single \cPZ-boson production.

Since the \cPZ\cPZ\ pair is produced in the collision of two hadrons,
the event might have jets from initial-state radiation.
We use jets reconstructed from particle-flow (PF)
candidates,
using the anti-$k_{\rm T}$ algorithm~\cite{Cacciari:2008gp} with a
distance parameter of 0.5. The jet transverse energy is corrected
using the CMS standard prescriptions for jet energy scale (JES)
calibration~\cite{ar:JER}.
Only jets with a corrected \pt greater than 10\GeV and reconstructed
within $\abs{\eta}<5$ are used in this analysis. Further corrections are
applied to reduce the effect of secondary proton-proton collisions
overlapping with the primary interaction (pileup).
An extra correction is applied to jets in the MC samples
to match the resolution observed in data.
In order to reject jets dominated by instrumental and beam-related
noise,
loose identification criteria are applied, based on the multiplicity
and energy fraction of charged and neutral particles.

{\tolerance=400
In order to suppress background coming from top quarks,
events are vetoed if they have a jet identified as a \PQb-quark jet (\PQb-tagged).
A requirement based on a combined secondary vertex
discriminator~\cite{Chatrchyan:2012jua}
is applied to \PQb-tagged jets with $\pt>20\GeV$ within
the tracker fiducial region ($\abs{\eta}<2.4$).
The misidentification probability for light-parton jets is about 10\%,
whereas the efficiency for \PQb-jets is more than 80\%.
To further reduce top-quark and other backgrounds with hadronic
activity, events are rejected if they contain any jet with
$\pt>30\GeV$.
\par}

A good \MET measurement is critical for the
extraction of the $\cPZ\cPZ\to2\Pl2\PGn$ signal given that the \MET
distinguishes this process from the DY background.
Since the average \MET of the signal is moderate ($\sim$50\GeV),
we cannot simply require a high \MET.
We follow the approach of constructing a ``reduced \MET'' variable,
as done in the D0~\cite{Abazov:2008yf,Abazov:2012cj} and
OPAL~\cite{Ackerstaff:1997rc} experiments.
The concept behind a reduced \MET is to reduce the
instrumental contribution to mismeasured \MET by considering possible
contributions to fake \MET.
In each event,
$\mathbf{\pt^{\text{miss}}}$ and jet momenta are decomposed along
an orthogonal set of axes in the transverse plane of the detector. One
of the axes is defined by the $\mathbf{\pt}$ of the charged
dilepton system,
the other perpendicular to it.
We define the recoil of the $\Pl^+\Pl^-$ system in two different ways:
(1) the clustered recoil ($\mathbf{R_{\mathrm{c}}}$) is the vectorial sum
of the momenta of the PF jets reconstructed in the event, and
(2) the unclustered recoil ($\mathbf{R_{\mathrm{u}}}$) is the vectorial
sum of the transverse momenta of all PF candidates in the event, with
the exception of the two leptons.
On each axis ($i=$ parallel/orthogonal to the dilepton system
$\mathbf{\pt}$), the reduced \MET projection is defined as
\begin{linenomath*}
\begin{equation*}
\text{reduced~\MET}^i = -\pt^{\Pl\Pl,i} - R_{\mathrm{c/u}}^i\,,
\end{equation*}
\end{linenomath*}
where $R_{\mathrm{c/u}}^i$ represents the choice of $R_{\mathrm{c}}$ or
$R_{\mathrm{u}}$ that
minimizes the absolute value of that reduced \MET component,
and $\pt^{\Pl\Pl,i}$ is a projection of the transverse momentum of the
\cPZ\ boson.
The presence of genuine \MET in the recoil of the charged dilepton system
is expected to be evident in the parallel projection, while the
component perpendicular to the $\Pl^{+}\Pl^{-}$ system is mostly
dominated by jet and \MET resolution. The absolute reduced \MET
variable is the sum in quadrature of the two components.
The reduced \MET shows better DY background suppression than the
standard PF \MET at the same signal efficiency.
It is also found to be more stable than the PF \MET under
variations in pileup conditions and JES.

The \MET balance variable is defined as the ratio between the PF \MET
and the transverse momentum of the leptonically decaying \cPZ\ boson,
namely $\MET/\pt^{\Pl\Pl}$.
Values of this variable far from unity identify events in which the
leptonic \cPZ-boson candidate is not well
balanced by genuine \MET from neutrinos, but recoils against
mismeasured jets or leptons.
The selected sample can still be contaminated by events with jets
with \pt below the veto threshold.

A mismeasurement of the jet energy can produce
mismeasured $\mathbf{\pt^{\text{miss}}}$
aligned with the jet direction in the transverse
plane. These events are characterized by a small azimuthal
angle between the $\mathbf{\pt^{\text{miss}}}$
vector and the closest jet,
$\Delta\phi$($\mathbf{\pt^{\text{miss}}}$,~jet).
This distribution is used to reject \cPZ+jets events that have a
small $\Delta\phi$ angle.
The mismeasurement of a lepton \pt can also
produce mismeasured \MET. Although this effect is usually
negligible, given the good lepton momentum resolution in CMS, events are
found where a large \MET value ($>$60\GeV) is accompanied by a small
angle between the $\mathbf{\pt^{\text{miss}}}$
and the $\mathbf{\pt}$ of a lepton. Events with
$\MET>60\GeV$ and
$\Delta\phi(\mathbf{\pt^{\text{miss}}}, \text{lepton})<0.2$\,rad
are therefore rejected.

In order to suppress the \PW\cPZ\ background, with both bosons decaying
leptonically, events are required to have no additional leptons.
To improve the rejection power, the \pt threshold is lowered to 3\GeV for
additional muons, and 10\GeV for
electrons. Furthermore, these muons and electrons are selected with
looser criteria than those used to reconstruct the \cPZ-boson candidate.

The variables described above are used to extract
the signal sample for the cross section measurement.
We optimize the requirements in the final selection in order to
minimize the total uncertainty in the measured cross section at
8\TeV (see Section~\ref{sec:xsec}). The same selection is applied to
the 7\TeV data. For this purpose, we scan a series of possible
analysis selections, in which
we vary the dilepton mass window and \pt threshold, the minimum \pt of jets used in the computation of
the reduced \MET variable, and the reduced \MET requirement.
We optimize the selection using MC estimates of the background
processes, or using predictions based on control samples in data from the DY, top-quark, and
\PW\PW\ backgrounds, as described in Section~\ref{sec:background}, and we
find similar results for the optimal requirements and for the measured
cross section.
For the final optimization we choose the selection obtained
using background estimates from data.
The requirements are summarized in Table~\ref{tab:selectioncuts}.
With this selection, the acceptance for
$\cPZ\cPZ\to 2\Pe2\PGn$ and $\cPZ\cPZ\to 2\Pgm2\PGn$ events is about 10\% for both channels, at 7 and 8\TeV.

\begin{table*}[htb]
 \centering
 \topcaption{Summary of the optimal signal selection.}
 \label{tab:selectioncuts}
 \renewcommand{\arraystretch}{1.2}
 \begin{tabular}{ll}
   \hline
   Variable  & Value \\
   \hline
   Dilepton invariant mass & $\abs{m(\Pl\Pl)-91} < 7.5$\GeV \\

   Dilepton \pt              & $\pt^{\Pl\Pl} > 45$\GeV \\
   \PQb-tagged jets          & Based on vertex info (for jet with $\pt > 20$\GeV) \\
   Jet veto                  & No jets with $\pt > 30\GeV$\\
   Reduced \MET              & $>$65\GeV \\
   \MET balance              & $0.4 < \MET/\pt^{\Pl\Pl} < 1.8$\\
   $\Delta\phi$($\mathbf{\pt^{\text{miss}}}$, jet)    & $>$0.5 rad \\
   $\Delta\phi$($\mathbf{\pt^{\text{miss}}}$, lepton)  & $>$0.2 rad \\
   Lepton veto               & No additional leptons (\Pe/\Pgm) with $\pt>10/3\GeV$ \\
   \hline
 \end{tabular}

\end{table*}

\section{Background estimation}
\label{sec:background}

Although the DY process does not include genuine \MET from neutrinos,
the tail of the reduced \MET distribution can be contaminated by these
events due to detector energy resolution, jet energy mismeasurements,
pileup energy fluctuations, and instrumental noise.
Given that the simulation may not fully reproduce detector and pileup
effects on the reduced \MET distribution, especially in the tails,
and that the simulation is limited in statistical precision,
we build a model of DY background from control samples in data.
For this purpose we use a process
that has similar jet multiplicity, underlying event, and pileup
conditions as the DY process for the region of interest at high boson
\pt: the production of prompt isolated photons in
association with jets (\cPgg\ +~jets)~\cite{PhysRevD.88.112009}.
We expect that an accurate description of the \MET distribution
and other related kinematic variables can be obtained from this photon
+ jets sample.
However, some corrections must be applied to the photon + jets sample to
ensure a good modeling of the DY process.
The yield of photon events is scaled to the observed charged dilepton
system yield
as a function of the boson \pt after applying
the jet veto to both samples.
This accounts for the differences in the selection efficiency of the
dilepton and photon candidates and corrects for the
trigger prescales, which are applied to the low-\pt photon triggers.

Only photons in the barrel region are used because the purity and
resolution are better than in other regions.
Following Ref.~\cite{Pedro:phot}, the selection of photon events is based
on shower shape, isolation in the tracker, and energy deposits in
ECAL, and HCAL.
After this selection, several processes with instrumental
\MET contribute to the photon sample: single \cPgg\ events, double
\cPgg\ events where one photon escapes detection or fails the
identification, and QCD events with a
jet misidentified as a photon. Processes
with genuine \MET can also contaminate this sample: \PW/\cPZ+\cPgg\
with the \PW/\cPZ\ boson decaying to $\Pl\PGn$/$\PGn\PGn$, or \PW+jets with the
\PW\ boson decaying to \Pe\PGn\ and
the electron misreconstructed as a photon.
Although these processes have generally lower
cross sections, they are characterized by large \MET values,
and thus contribute to the tails of the distribution, where it is most
important to measure the residual instrumental background.
In order to reduce these background contributions, specific selections are applied.
The event must have exactly one photon and no leptons.
Only jets with
$\Delta R = \sqrt{\smash[b]{\left(\Delta\phi\right)^2 +
\left(\Delta\eta\right)^2}} > 0.4$
from the photon are used for all the
jet-related selections (jet veto, reduced \MET, \etc).
To avoid misreconstruction of the photon energy, a conversion veto
is applied using the number of missing expected tracker hits and
the distance of closest approach between the reconstructed conversion tracks.

The remaining contribution from $\PW+\cPgg$ and $\PW/\cPZ+\cPgg$ events after
this selection is estimated from simulation and subtracted from the photon
data model. For this purpose, a set of simulated photon samples is used
that includes \cPgg~+~jets, QCD events with a jet misidentified as a photon (generated with \PYTHIA),
$\PW+\cPgg \to \Pl\PGn\cPgg$, and $\cPZ+\cPgg \to \PGn\PGn\cPgg$ (generated with \MADGRAPH).
These samples are normalized to their respective cross sections computed at NLO in QCD.
The full set of MC samples is reweighted and corrected following the
same procedure as that used for the photon data sample.
Finally, the photon data are corrected as a function of \MET by
multiplying them by unity minus the fraction of electroweak processes in
the simulation.

We apply a different data-based method to estimate the total number of
background events from processes that do not involve a \cPZ\ boson:
\ie \PW\PW\ and top-quark production.
We denote these events as
nonresonant background (NRB). In order to measure this contribution, a
control sample based on \Pe\Pgm\ candidate events
is selected by applying the same requirements as in the main analysis.
The NRB yields in the same-flavor channels (\Pe\Pe\ and \Pgm\Pgm)
are obtained by scaling the number of events in the control sample.
The rescaling is done by means of correction factors, measured from
the sidebands (SB) of the \cPZ-boson mass peak, \ie
in the regions 55--70 and 110--200\GeV.
The scale factors are measured in a looser selection region in order
to improve the statistical precision.
We require the reduced $\MET > 65$\GeV in order to suppress the DY
contribution from $\Pgt^+\Pgt^-$.
We also require at least one \PQb-tagged jet with $\pt>20\GeV$, to
further reduce DY and other backgrounds, and increase the fraction of
top-quark events.
The scale factors are defined as follows:
\begin{linenomath*}
\begin{equation}
\alpha_{\Pe\Pe/\Pgm\Pgm} = N^\mathrm{SB}_{\Pe\Pe/\Pgm\Pgm} / N^\mathrm{SB}_{\Pe\Pgm},
\label{eq:alphadef}
\end{equation}
\end{linenomath*}
and the NRB contamination in the \cPZ-peak region is:
\begin{linenomath*}
\begin{equation}
N^\text{peak}_{\Pe\Pe/\Pgm\Pgm} = \alpha_{\Pe\Pe/\Pgm\Pgm}\, N^\text{peak}_{\Pe\Pgm}.
\label{eq:NRBest}
\end{equation}
\end{linenomath*}
The validity of the method is tested in simulation by comparing the
predicted background to the expected number of \PW\PW\ and top-quark events.

\begin{figure}[tbp]
\centering \includegraphics[width=0.48\textwidth]{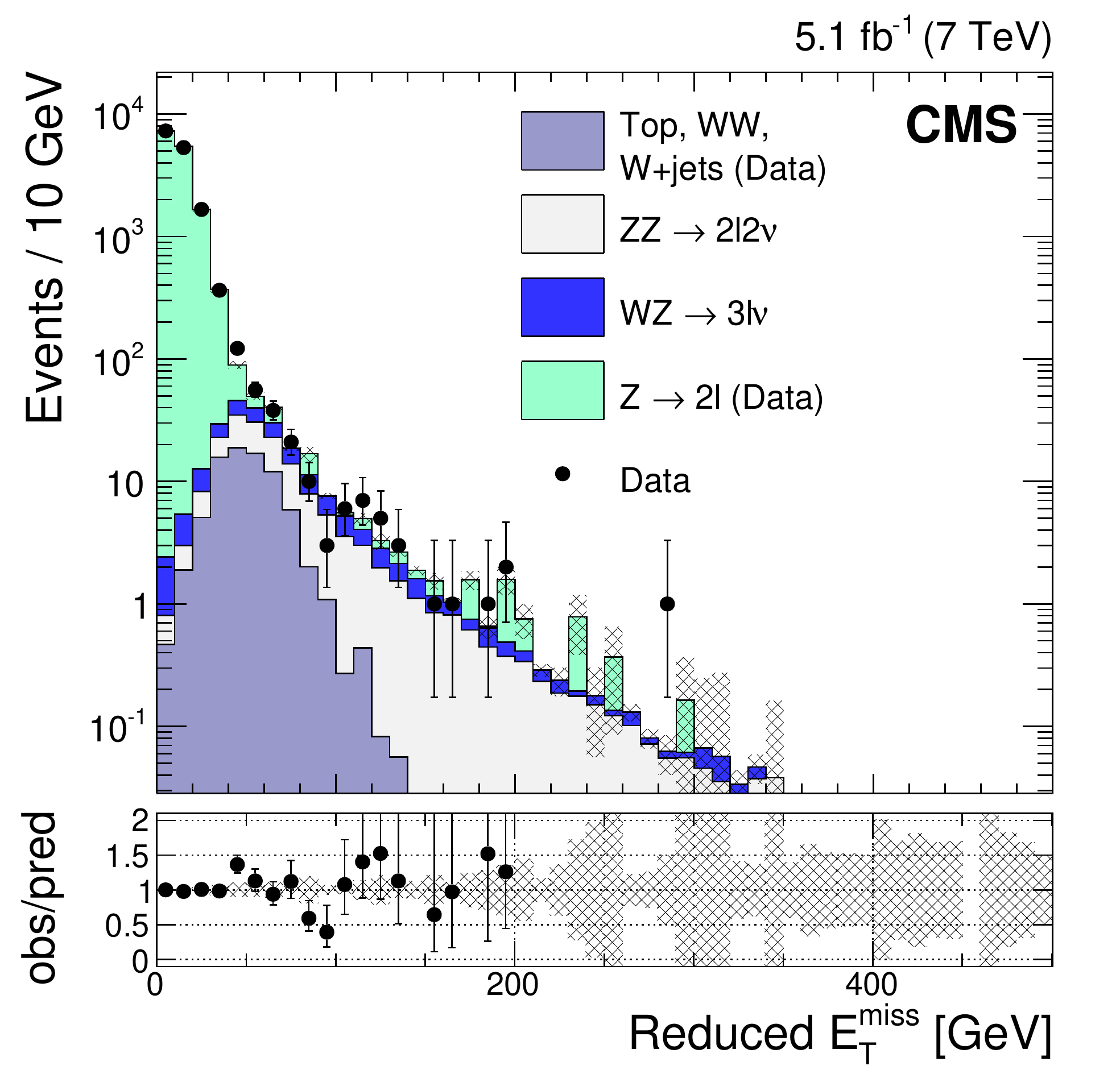}
\centering \includegraphics[width=0.48\textwidth]{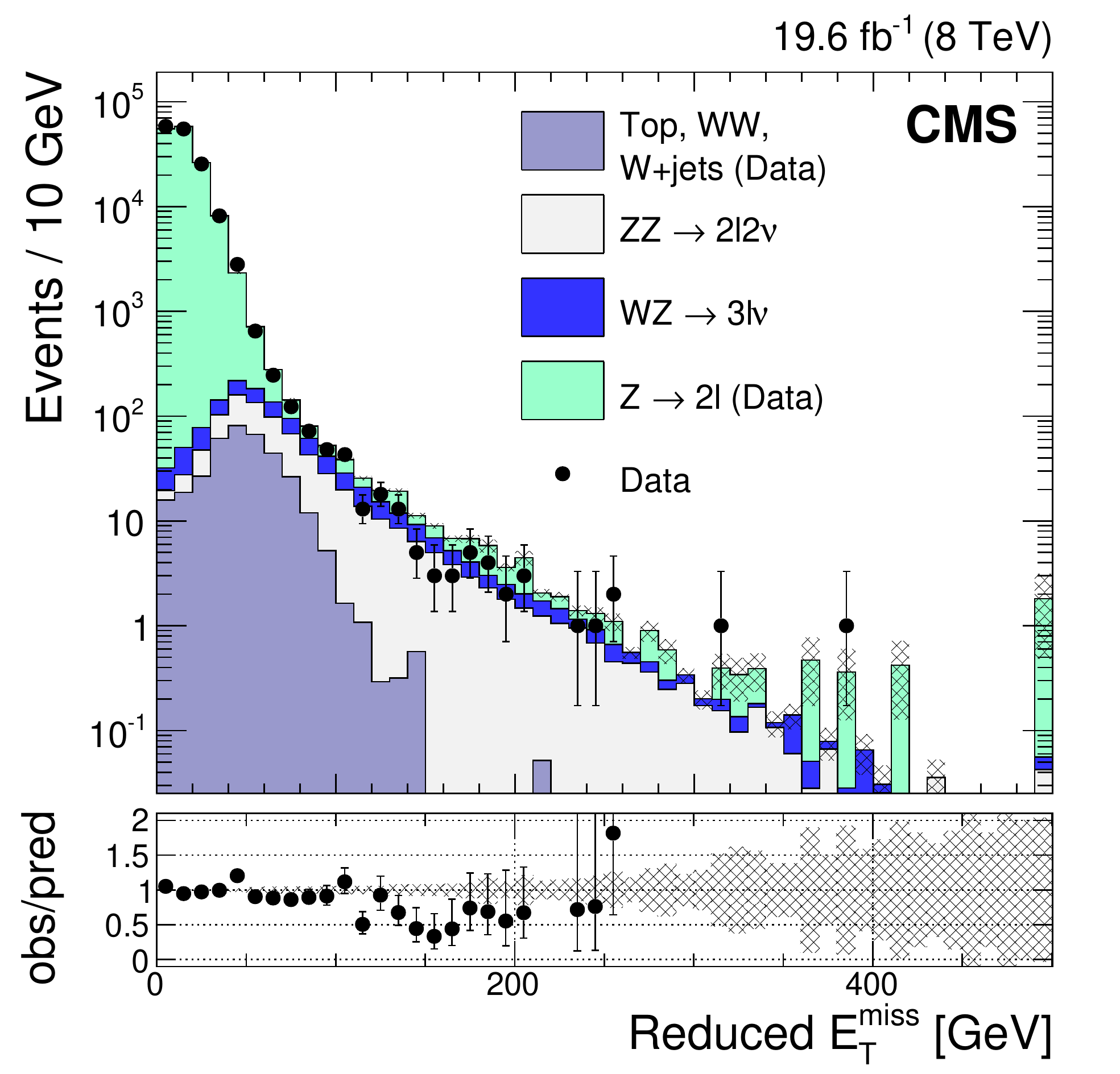}
\caption{Reduced \MET spectrum in the inclusive $\Pl\Pl$ ($\Pl =
  \Pe, \Pgm$) channel at 7\TeV (\cmsLeft) and 8\TeV (\cmsRight),
  using the photon model to describe the DY contribution and NRB
  modeling for \PW\PW, \PW\ + jets,
  and top-quark production, after selections on the dilepton invariant mass and \pt,
  jet veto, \PQb-tagged jet veto, third lepton veto, and $\Delta\phi$($\mathbf{\pt^{\text{miss}}}$,~jet),
  as described in Section~\ref{sec:selection}.
  The gray error band represents the statistical
  uncertainty in the predicted yields.}
\label{fig:DY_Met_final}
\end{figure}

Figure~\ref{fig:DY_Met_final} shows the reduced \MET
distributions in dilepton data and simulation, using the photon model
to describe the DY background
and the data-driven estimation for NRB.
A good agreement is found
in the region dominated by the DY process, up to about 80\GeV, while
the higher part of the spectrum is dominated by diboson production.
The error bands shown in Fig.~\ref{fig:DY_Met_final} represent the
statistical uncertainty in the predicted yields.
A systematic uncertainty in the final DY event yield estimated with
this method is computed as the relative difference between dilepton
yields in data and simulation, in a control region
with $\MET<60\GeV$,
and it has been found to be 25\%\,(40\%) at 7\,(8)\TeV. This systematic
uncertainty is not shown in Figure~\ref{fig:DY_Met_final}.

\section{Systematic uncertainties}
\label{sec:systematics}

Different sources of systematic uncertainty are associated with the
expected yields and distributions of signal and background processes
and of the data.
The uncertainties reported in the following paragraphs affect
the final event yields of the relevant processes.

\paragraph{Statistical uncertainty of the simulated and control samples.}
For the processes estimated from simulation, \cPZ\cPZ\ and \PW\cPZ, the
limited size of the MC sample
affects the precision of the modeling, and
is therefore taken as a systematic uncertainty in the shape of the
kinematic distributions used in the cross section measurement and
ATGC limit setting.
Similarly, the backgrounds estimated from data are limited by the
size of the control samples described in
Section~\ref{sec:background}: the
\Pe\Pgm\ sample for nonresonant backgrounds and the
\cPgg+jets sample for DY background.
These uncertainties are treated in the same way as those backgrounds
that are estimated from simulation. This systematic uncertainty has
been computed in different reduced \MET bins or different \pt bins and
is used as shape errors in the fit.

\paragraph{Cross sections of \texorpdfstring{\cPZ\cPZ}{ZZ} and \texorpdfstring{\PW\cPZ}{WZ}.}
The cross sections for $\Pp\Pp \to \cPZ\cPZ+\mathrm{X}\to 2\Pl2\PGn+\mathrm{X}$
and $\Pp\Pp\to\PW\cPZ+\mathrm{X}\to 3\Pl\PGn+\mathrm{X}$
processes are calculated using
\MCFM version 6.2~\cite{bib:MCFM}, and using PDFs from
the Les Houches accord PDF (\textsc{lhapdf}) program, version
5.8.7~\cite{Whalley:2005nh}.
The PDF+$\alpha_\mathrm{S}$ uncertainty in the \PW\cPZ\ cross section
is evaluated as the
maximum spread of the cross sections computed at $\mu_R = \mu_F = m_\Z$
with three PDF sets, including the corresponding uncertainties
from one standard deviation
variation of the PDF parameters and the $\alpha_\mathrm{S}$
value~\cite{Botje:2011sn}.
It is found to be
3.1\%\,(4.2\%) at 7\,(8)\TeV.

The uncertainty from the renormalization and factorization scales is
evaluated as the maximum difference between the central value of the
cross section at $\mu_R = \mu_F = m_\cPZ$ and the central values
computed at $\mu_R = \mu_F = m_\cPZ/2$ and $2\,m_\cPZ$, using
each of the three PDFs recommended in Ref.~\cite{Botje:2011sn}.
An uncertainty of 5.9\%\,(5.4\%) at 7\,(8)\TeV is found for the
\PW\cPZ\ background. For the \cPZ\cPZ\ signal, we evaluate this theoretical
uncertainty in the case of the exclusive production with 0 jets, to
take into account the jet-veto applied in the signal selection,
following the prescription described in
Refs.~\cite{Stewart:2011cf,an:N2011005}.
The exclusive cross section for \cPZ\cPZ+0~jets is
\mbox{$\sigma_{0j} = \sigma_{\geq 0j} - \sigma_{\geq 1j}$},
where $\sigma_{\geq nj}$
is the inclusive cross section of \cPZ\cPZ\ $+$ at least $n$ jets, where $n=0,1$.
According to Ref.~\cite{Stewart:2011cf}, $\sigma_{\geq 0j}$ and
$\sigma_{\geq 1j}$ are essentially uncorrelated, thus the uncertainty in
$\sigma_{0j}$ can be computed as
$\epsilon_{0j} = \sqrt{\smash[b]{\epsilon_{\geq 0j}^2 + \epsilon_{\geq 1j}^2}}$,
where
$\epsilon_{\geq 0j}$ and $\epsilon_{\geq 1j}$ are the uncertainties in
$\sigma_{\geq 0j}$ and $\sigma_{\geq 1j}$, respectively.
The cross sections are computed with \MCFM,
including the acceptance requirements on
lepton \pt and $\eta$, charged dilepton mass, and \MET, as well as
the jet veto, when relevant.
The cross section uncertainties are estimated by varying
the renormalization and factorization scales, as explained above.
Since the charged dilepton \pt spectrum is the observable from which
limits on ATGCs are derived, the uncertainty in $\sigma_{0j}$ is
computed in different intervals of charged dilepton \pt.

The uncertainty in the NLO EW correction to \cPZ\cPZ\ production,
corresponding to missing higher-order terms in the computation, is
estimated as the product of the NLO QCD and EW
corrections~\cite{Bierweiler:2013dja}. The uncertainty in the EW
correction to \PW\cPZ\ production is estimated as 100\% of the
correction, to account for the poorly known fraction of
photon+quark-induced events~\cite{Baglio:2013toa} passing the jet
veto.

\paragraph{Acceptance.}
The kinematic acceptance for the signal is computed using \MCFM.
Kinematic requirements, based on those used in the signal selection,
are applied to the charged leptons and neutrinos at the generator
level. The acceptance is determined by comparing the cross sections
with and without the kinematic requirements.
The systematic uncertainty is evaluated as the variation in the
acceptance resulting from varying the renormalization
and factorization scales from $m_\cPZ$ to $m_\cPZ/2$ and $2\,m_\cPZ$, summed in
quadrature with the variation obtained from using different PDF sets
and from varying the PDF parameters and the $\alpha_\mathrm{S}$ value
by one standard deviation.
The result is 2.8\% at both 7 and 8\TeV.

\paragraph{Luminosity.}
The uncertainty in the luminosity measurement is
2.2\% in 2011, and 2.6\% in 2012~\cite{CMS-PAS-LUM-13-001}.

\paragraph{Lepton trigger and identification efficiency.}
Lepton trigger and identification efficiencies are determined from data,
using the tag-and-probe technique with $\cPZ\ \to \Pl\Pl$
events~\cite{First:WZ}, and
used to correct the simulated samples. The total
uncertainty in the lepton efficiency
amounts to about 3\%
for \Pe\Pe\ events, and 4\% for \Pgm\Pgm\ events.

\paragraph{Lepton momentum scale.}
The systematic uncertainty in the lepton momentum scale is computed
by shifting the nominal momenta by $\pm 1\sigma$ and propagating the
variations to the reduced \MET. We assume an uncertainty of 2\%
(3.5\%) in the energy of electrons reconstructed in the ECAL barrel
(endcap),
and 1\% in the muon
momentum.
The resulting variations of the final
yields are
2.5\% for the \Pe\Pe\  channel, and 1.0\% for
the \Pgm\Pgm\ channel and they are treated as a shape uncertainty.

\paragraph{Jet energy scale and resolution.}
The uncertainty in the calibration of the jet energy scale
directly affects the jet veto,
the calculation of reduced \MET, and the selection of the balance
variable.
The JES uncertainty is estimated by shifting the jet
energies by $\pm 1\sigma$ and propagating the variations to the
reduced \MET and all the other relevant observables.
Uncertainties in the final yields of 3--4\,(7--8)\% are found for both
the \Pe\Pe\ and \Pgm\Pgm\ final states at 7\,(8)\TeV.

Similarly, a systematic uncertainty in jet energy resolution (JER) is
computed. As explained above, the energy
of jets in simulation is corrected
to reproduce the resolution observed in
data. Such corrections are varied according to their
uncertainties and these variations are propagated to all the
observables and selections dependent on jet energy.
An uncertainty in the final yields of less than 1\% is found in both
\Pe\Pe\ and \Pgm\Pgm\ final states: 0.4\%\,(0.8\%) at 7\,(8)\TeV.

Since the shapes of the distributions are expected to be affected
by variations in the JES and the JER, these sources are
treated as shape uncertainties in the extraction of the cross section.

\paragraph{\texorpdfstring{\PQb}{b}-jet veto.}
The \PQb-tagging efficiency is taken from Ref.~\cite{CMS-DP-2013-005}.
In simulation, the nominal working point
for this \PQb-tagger is shifted to reproduce the
efficiency observed in data. The uncertainty in the measured
efficiency is propagated to the event
yields of the processes estimated from simulation
by applying further shifts to the discriminator
threshold.
A very small uncertainty in the final yields of the MC
samples is found: 0.1--0.15\% at both 7 and 8\TeV.

\paragraph{Pileup.}
Simulated samples are reweighted to reproduce
the pileup conditions observed in data.
To compute the uncertainty related to this procedure, we
shift
the number of interactions by
8\% when reweighting the simulated samples.
The variation of the final yields induced by
this procedure is less than 1\% in \cPZ\cPZ\ and
\PW\cPZ\ processes. However, the shapes of the kinematic distributions
can vary in this procedure, so the varied distributions are used
as shape uncertainties in the cross section fit.

\paragraph{Drell--Yan.}
The uncertainty in the DY contribution is propagated from the
uncertainty in the reweighted photon spectrum that is
used in the estimate of DY background from data, and is dominated
by the subtraction of backgrounds due to EW processes.
As explained in Section~\ref{sec:background}, the DY background estimate is
assigned an uncertainty of 25\%\,(40\%) at 7\,(8)\TeV,
evaluated from the relative difference between dilepton
yields in data and simulation in a control region.

\paragraph{Top-quark and \texorpdfstring{\PW\PW}{WW} backgrounds.}
The uncertainty in the estimate of the NRB
is derived from
the statistical uncertainties in the scale factors in Eq.~(\ref{eq:alphadef}),
and from a closure test
of the data-driven method for the measurement of this background
performed on simulated data.
It is found to be about 20\% at both 7 and 8\TeV.

\section{Measurement of the \texorpdfstring{\cPZ\cPZ}{ZZ} production cross section}
\label{sec:xsec}

\begin{figure}[!Hhtb]
  \centering
  \includegraphics[width=0.48\textwidth]{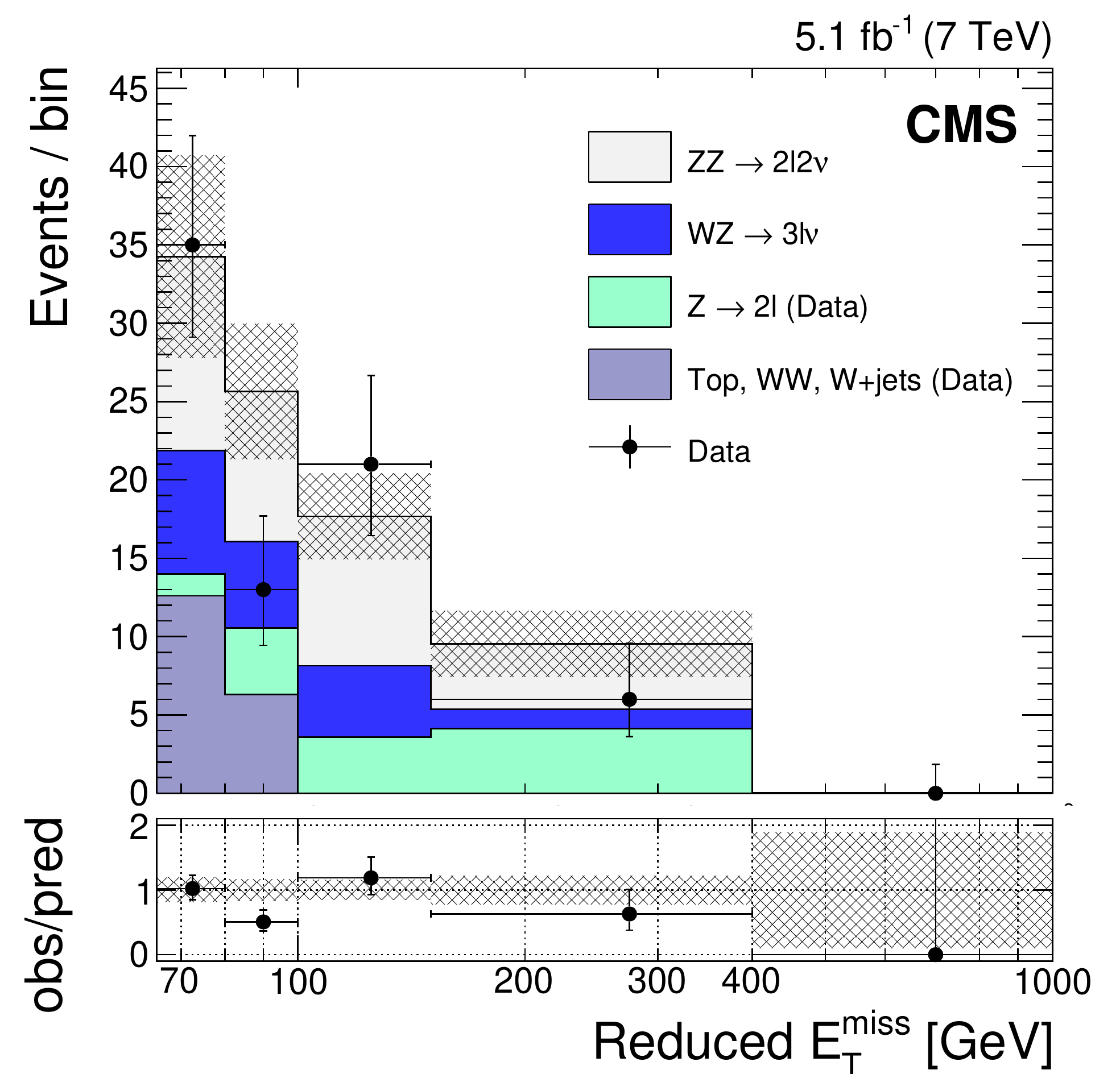}
  \includegraphics[width=0.48\textwidth]{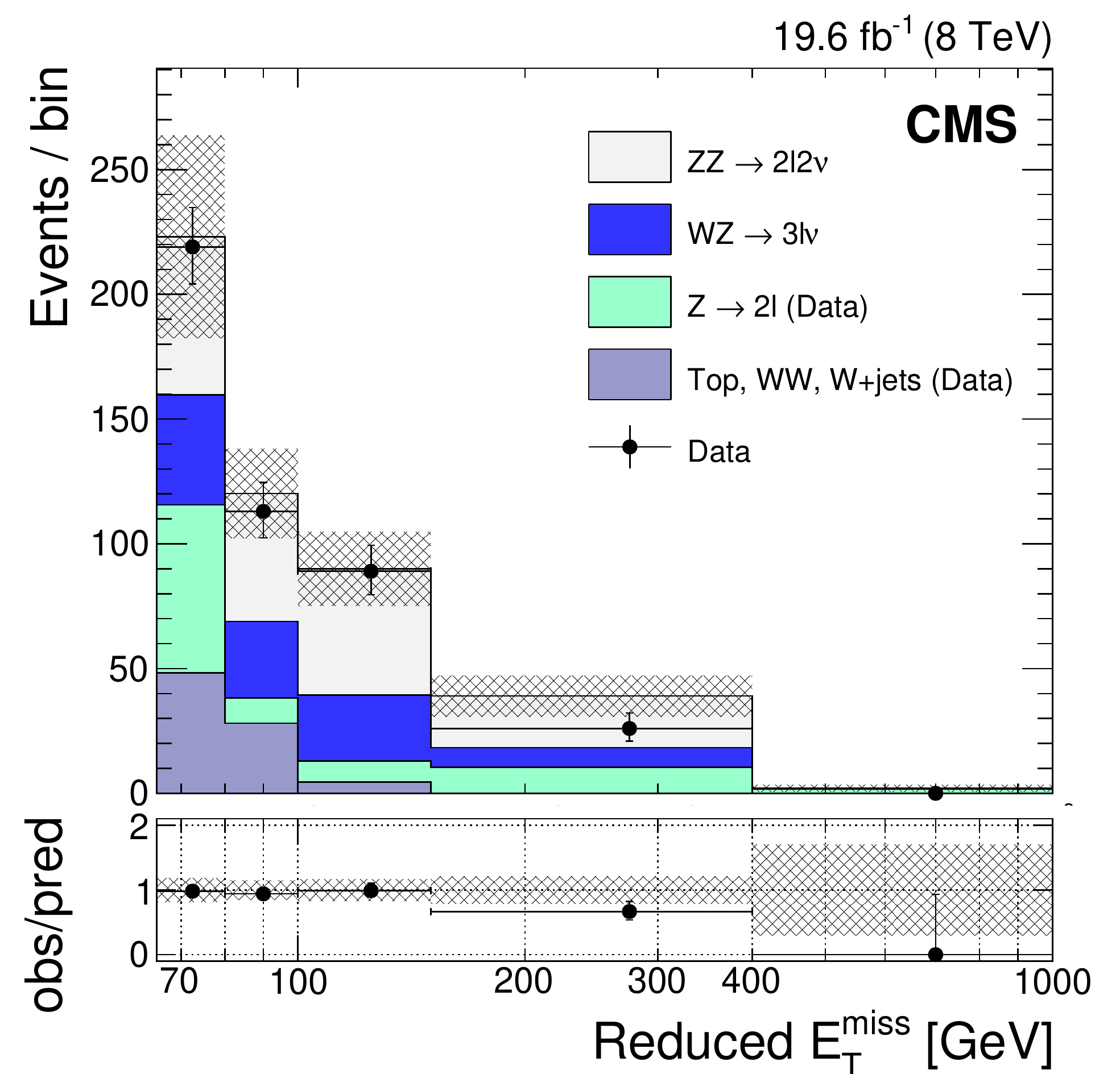}
  \caption{Reduced \MET distribution in $\Pl\Pl$ ($\Pl = \Pe,\Pgm$)
    channels, after the full selection, at 7\TeV (\cmsLeft) and
    8\TeV (\cmsRight). The DY and \PW\PW, \PW+jets, and top
    backgrounds are estimated with data-driven methods. The gray error
    band includes statistical and systematic uncertainties in the
    predicted yields. In the bottom plots, vertical error bars and
    bands are relative to the total predicted yields. In all plots,
    horizontal error bars indicate the bin width.}
  \label{fig:redMet_final_gamma}
\end{figure}

We extract the \cPZ\cPZ\ production cross section using a profile
likelihood fit~\cite{PhysRevD.86.010001} to the reduced-\MET distribution, shown in
Fig.~\ref{fig:redMet_final_gamma}. The fit takes into account the
expectations for the different background processes and the \cPZ\cPZ\
signal.
Each systematic uncertainty is introduced in the fit as a
nuisance parameter with a log-normal prior. For the signal we consider a
further multiplicative factor, which is the ratio of the cross section
measured in data to the expected theoretical value, \ie the
signal strength $\mu=\sigma/\sigma_\text{th}$. Maximizing
the profile likelihood, we obtain the \cPZ\cPZ\ production cross
section from the signal strength parameter, as well as optimal fits of
the background yields by varying nuisance parameters within their
constraints. Table~\ref{tab:prepostfits} shows the expected signal and
background yields, and the corresponding values
after the combined fit to the \Pe\Pe\ and \Pgm\Pgm\ channels.
The uncertainties include both the statistical and systematic components.

\begin{table*}[htp]
\centering
\topcaption{Predicted signal and background yields at 7 and 8\TeV, and
  corresponding values obtained from the combined maximum likelihood
  fit to the \Pe\Pe\ and \Pgm\Pgm\ channels.
  The uncertainties include both the statistical and systematic
  components.}
\label{tab:prepostfits}
\begin{tabular}{cccccc}
  \hline
  Dataset & Process & Channel & Predicted yield & Fitted yield & Observed \\
  \hline
  \multirow{10}{*}{7\TeV}& \multirow{2}{*}{$\cPZ\cPZ\to2\Pl2\PGn$}& \Pe\Pe\     & $14.0 \pm 1.9$ & $12.0 \pm 4.4$ & $-$\\
                         &                                        & \Pgm\Pgm\ & $21.7 \pm 3.2$ & $18.4 \pm 6.8$ & $-$\\
                         \cline{2-6}
                         & \multirow{2}{*}{$\PW\cPZ\to3\Pl\PGn$}  & \Pe\Pe\     & $7.7  \pm 0.9$ & $7.9  \pm 1.0$ & $-$\\
                         &                                        & \Pgm\Pgm\ & $11.5 \pm 1.6$ & $11.6 \pm 1.2$ & $-$\\
                         \cline{2-6}
                         & \multirow{2}{*}{\cPZ\ + jets}          & \Pe\Pe\     & $5.0  \pm 2.7$ & $4.8  \pm 2.3$ & $-$\\
                         &                                        & \Pgm\Pgm\ & $8.3  \pm 4.8$ & $4.8  \pm 3.0$ & $-$\\
                         \cline{2-6}
                         & \multirow{2}{*}{Nonresonant}           & \Pe\Pe\     & $7.7  \pm 3.1$ & $7.4  \pm 2.3$ & $-$\\
                         &                                        & \Pgm\Pgm\ & $11.2 \pm 4.8$ & $9.2  \pm 3.1$ & $-$\\
                         \cline{2-6}
                         & \multirow{2}{*}{Total}                 & \Pe\Pe\     & $34.4 \pm 6.2$ & $32.1 \pm 3.9$ & $35$ \\
                         &                                        & \Pgm\Pgm\ & $52.7 \pm 9.7$ & $44.0 \pm 5.3$ & $40$ \\
  \hline
  \hline
  \multirow{10}{*}{8\TeV}& \multirow{2}{*}{$\cPZ\cPZ\to2\Pl2\PGn$}& \Pe\Pe\     & $77  \pm 16$    & $69    \pm 13 $ & $-$\\
                         &                                        & \Pgm\Pgm\ & $109 \pm 23$    & $100   \pm 19 $ & $-$\\
                         \cline{2-6}
                         & \multirow{2}{*}{$\PW\cPZ\to3\Pl\PGn$}  & \Pe\Pe\     & $45  \pm  6$    & $43.9  \pm 5.6$ & $-$\\
                         &                                        & \Pgm\Pgm\ & $64  \pm  8$    & $63.8  \pm 7.3$ & $-$\\
                         \cline{2-6}
                         & \multirow{2}{*}{\cPZ\ + jets}          & \Pe\Pe\     & $36  \pm 12$    & $27.7  \pm 7.9$ & $-$\\
                         &                                        & \Pgm\Pgm\ & $63  \pm 21$    & $52    \pm 14 $ & $-$\\
                         \cline{2-6}
                         & \multirow{2}{*}{Nonresonant}           & \Pe\Pe\     & $31  \pm  9$    & $34.1  \pm 7.2$ & $-$\\
                         &                                        & \Pgm\Pgm\ & $50  \pm 14$    & $54    \pm 12 $ & $-$\\
                         \cline{2-6}
                         & \multirow{2}{*}{Total}                 & \Pe\Pe\     & $189 \pm 31$    & $174.7 \pm 10$ & $176$\\
                         &                                        & \Pgm\Pgm\ & $286 \pm 49$    & $269.8 \pm 15$ & $271$\\
  \hline
\end{tabular}
\end{table*}

The cross sections are extracted from individual fits to the \Pe\Pe\ and
\Pgm\Pgm\ channels and from a simultaneous fit to both channels.
Table~\ref{tab:xseccomb} reports the measured
$\Pp\Pp \to \cPZ\cPZ \to 2\Pl2\PGn$ exclusive cross section, \ie the
production cross section of \cPZ\cPZ\ pairs with mass
$60<M_\cPZ< 120\GeV$,
with no restrictions on lepton acceptance nor jet number,
times the branching fraction to final states with
two charged leptons of a given flavor and
two neutrinos of any flavor.
This is obtained by rescaling the theoretical
prediction for the exclusive cross section in the same kinematic range
by the fitted signal strength. These theoretical predictions are
computed at NLO in QCD with \MCFM and corrected for NLO EW
effects:
$79^{+4}_{-3}$ ($97^{+4}_{-3}$)\unit{fb} at 7\,(8)\TeV.

\begin{table*}[htp]
\centering
\topcaption{Cross sections (fb) for process $\Pp\Pp \to \cPZ\cPZ \to 2\Pl2\PGn$
  (where $\Pl$ denotes a charged lepton of a given flavor, $\PGn$ a
  neutrino of any flavor) at 7 and 8\TeV, with both \cPZ\ boson
  masses in the range 60-120\GeV, measured in the \Pe\Pe\ and
  \Pgm\Pgm\ channels and the two channels combined.}
\label{tab:xseccomb}
\begin{tabular}{cll}
  \hline
  \multirow{2}{*}{Channel} &
  \multicolumn{1}{c}{\multirow{2}{*}{$\sqrt{s} = 7\TeV$}} &
  \multicolumn{1}{c}{\multirow{2}{*}{$\sqrt{s} = 8\TeV$}} \\
  & & \\
  \hline
  \multirow{2}{*}{\Pe\Pe} &
  \multirow{2}{*}{$98_{-31}^{+35}\stat\,_{-22}^{+27}\syst\pm 2\lum$} &
  \multirow{2}{*}{$83_{-16}^{+17}\stat\,_{-19}^{+26}\syst\pm 2\lum$} \\
  & & \\
  \multirow{2}{*}{\Pgm\Pgm} &
  \multirow{2}{*}{$47_{-21}^{+24}\stat\,_{-19}^{+20}\syst\pm 1\lum$} &
  \multirow{2}{*}{$98_{-14}^{+14}\stat\,_{-22}^{+29}\syst\pm 3\lum$} \\
  & & \\
  \hline
  \multirow{2}{*}{Combined} &
  \multirow{2}{*}{$66_{-18}^{+20}\stat\,_{-14}^{+18}\syst\pm 1\lum$} &
  \multirow{2}{*}{$92_{-10}^{+11}\stat\,_{-19}^{+25}\syst\pm 2\lum$} \\
  & & \\
  \hline
  \multirow{2}{*}{Theory} & \multirow{2}{*}{$79^{+4}_{-3}\thy$} & \multirow{2}{*}{$97^{+4}_{-3}$\thy} \\
  &  &  \\
  \hline
\end{tabular}
\end{table*}

The measured inclusive \cPZ\cPZ\ cross section is obtained by rescaling the
theoretical inclusive cross section computed in the zero-width
approximation~\cite{bib:MCFM} and corrected for NLO EW
effects~\cite{Bierweiler:2013dja} (see
Section~\ref{sec:introduction}), by the same fitted signal
strength. This procedure properly accounts for the contribution of
virtual photon decays to the charged-lepton pair production, and
yields a measured cross section that can be compared directly with
theoretical calculations of inclusive pure \cPZ\cPZ\ production in the
zero-width approximation. The results are:
\ifthenelse{\boolean{cms@external}}{
\begin{multline*}
7\TeV:\\\sigma(\Pp\Pp \to \cPZ\cPZ) =
5.1_{-1.4}^{+1.5}\stat\,_{-1.1}^{+1.4}\syst\pm 0.1\lum\unit{pb},
\end{multline*}
\begin{multline*}
8\TeV:\\\sigma(\Pp\Pp \to \cPZ\cPZ) =
7.2_{-0.8}^{+0.8}\stat\,_{-1.5}^{+1.9}\syst\pm 0.2\lum\unit{pb}.
\end{multline*}
}{
\begin{linenomath*}
\begin{equation*}
7\TeV:\quad\sigma(\Pp\Pp \to \cPZ\cPZ) =
5.1_{-1.4}^{+1.5}\stat\,_{-1.1}^{+1.4}\syst\pm 0.1\lum\unit{pb},
\end{equation*}
\begin{equation*}
8\TeV:\quad\sigma(\Pp\Pp \to \cPZ\cPZ) =
7.2_{-0.8}^{+0.8}\stat\,_{-1.5}^{+1.9}\syst\pm 0.2\lum\unit{pb}.
\end{equation*}
\end{linenomath*}
}
This is the first cross section measurement in the $2\Pl2\PGn$ channel at 8 TeV.
The measurements are less than one standard
deviation from the SM predictions at both 7 and 8\TeV.
The uncertainties are approximately twice as large as
those from the CMS measurement in the
$4\ell$ channel~\cite{:2012rg,Khachatryan:2014dia}, and the channels
agree within uncertainties.

The $p$-values of the simultaneous fit to the \Pe\Pe\ and \Pgm\Pgm\
channels are 0.335\,(0.569) at 7\,(8)\TeV. The data are also
consistent with the reduced \MET spectra uncorrected for NLO EW effects,
but with slightly smaller $p$-values of 0.322\,(0.477) at 7\,(8)\TeV.
The application of EW corrections thus improves the modeling of the diboson
processes and leads to a better agreement between the simulated and observed spectra.

Table~\ref{tab:systemxs} shows a summary of the sources of
systematic uncertainty described in Section~\ref{sec:systematics},
with the corresponding contributions to the total systematic
uncertainty in the cross sections.

\begin{table*}[hbt]
  \centering
  \topcaption{Systematic uncertainties in the cross sections due to
    each source separately, after the maximum likelihood fit to
    extract the \cPZ\cPZ\ cross section. The uncertainties marked with an
    asterisk ($\ast$) are used as shape uncertainties in the fit.}
  \label{tab:systemxs}
  \begin{tabular}{lcc}
  \hline
  \multirow{2}{*}{Source of uncertainty}
  & \multicolumn{2}{c}{Uncertainty [\%]} \\
   & 7\TeV & 8\TeV \\
  \hline
  ($\ast$) MC statistics: \cPZ\cPZ\ (\Pe\Pe\ channel)        & 0.8  & 1.0   \\
  ($\ast$) MC statistics: \cPZ\cPZ\ (\Pgm\Pgm\ channel)      & 1.3  & 1.1   \\
  ($\ast$) MC statistics: \PW\cPZ\ (\Pe\Pe\ channel)         & 1.7  & 0.9   \\
  ($\ast$) MC statistics: \PW\cPZ\ (\Pgm\Pgm\ channel)       & 1.7  & 1.0   \\
  ($\ast$) Control sample statistics: DY (\Pe\Pe\ channel)   & 6.9  & 2.3   \\
  ($\ast$) Control sample statistics: DY (\Pgm\Pgm\ channel) & 5.8  & 4.9   \\
  ($\ast$) Control sample statistics: NRB (\Pe\Pe\ channel)  & 6.3  & 3.0   \\
  ($\ast$) Control sample statistics: NRB (\Pgm\Pgm\ channel)& 8.1  & 4.4   \\
  \PW\cPZ\ cross section: PDF+$\alpha_{\mathrm S}$           & 1.9  & 2.6   \\
  ($\ast$) \cPZ\cPZ\ + \PW\cPZ\ cross section: scales        & 17   & 16    \\
  ($\ast$) \cPZ\cPZ\ + \PW\cPZ\ cross section: NLO EW corr.  & 2.4  & 2.3   \\
  Signal acceptance                                          & 2.8  & 2.8   \\
  ($\ast$) Pileup                                            & 0.5  & 1.1   \\
  Muon trigger, ID, isolation                                & 4.1  & 3.6   \\
  Electron trigger, ID, isolation                            & 1.7  & 2.0   \\
  ($\ast$) Lepton momentum scale                             & 2.6  & 3.7   \\
  ($\ast$) JES                                               & 6.0  & 12    \\
  ($\ast$) JER                                               & 0.8  & 1.4   \\
  ($\ast$) Unclustered \MET                                  & 2.1  & 3.2   \\
  ($\ast$) \PQb-jet veto                                     & 0.3  & 0.5   \\
  Drell--Yan bkg. normalization                              & 6.6  & 8.4   \\
  Top-quark \& \PW\PW\ bkg. normalization                    & 7.7  & 7.1   \\
  \hline
  Total systematic uncertainty                               & 24.6 & 23.5  \\
  Statistical uncertainty                                    & 28.0 & 11.9  \\
  \hline

  \end{tabular}
\end{table*}%

\section{Anomalous couplings}
\label{sec:atgc}

The existence of neutral trilinear gauge couplings is forbidden at the
tree level, but allowed in some extensions of the SM~\cite{ar:aTGC}.
The \cPZ\cPZ\ production process provides a way to probe the existence
of such anomalous couplings at the \cPZ\cPZ\cPZ\ and
\cPgg\cPZ\cPZ\ vertices.

Neutral couplings $\mathrm{V}^{(\ast)}\cPZ\cPZ$ ($\mathrm{V}
= \cPZ, \cPgg$) can be described
using the following effective Lagran\-gian~\cite{Hagiwara:1987}:
\ifthenelse{\boolean{cms@external}}{
\begin{multline}
\label{eq:atgc_lagrangian}
\mathcal{L}_{\mathrm{V}\cPZ\cPZ}  = -\frac{e}{M_{\Z}^2} \Bigl\{
\left[f_4^\cPgg\left(\partial_\mu F^{\mu\alpha}\right)
+f_4^\cPZ\left(\partial_\mu Z^{\mu\alpha}\right)\right]
Z_\beta\left(\partial^\beta Z_\alpha\right)\\
-\left[f_5^\cPgg\left(\partial^\mu F_{\mu\alpha}\right)
+f_5^\cPZ\left(\partial^\mu Z_{\mu\alpha}\right)\right]
\tilde{Z}^{\alpha\beta}Z_\beta
\Bigr\},
\end{multline}
}{
\begin{linenomath*}
\begin{equation}
\label{eq:atgc_lagrangian}
\mathcal{L}_{\mathrm{V}\cPZ\cPZ}  = -\frac{e}{M_{\Z}^2} \left\{
\left[f_4^\cPgg\left(\partial_\mu F^{\mu\alpha}\right)
+f_4^\cPZ\left(\partial_\mu Z^{\mu\alpha}\right)\right]
Z_\beta\left(\partial^\beta Z_\alpha\right)
-\left[f_5^\cPgg\left(\partial^\mu F_{\mu\alpha}\right)
+f_5^\cPZ\left(\partial^\mu Z_{\mu\alpha}\right)\right]
\tilde{Z}^{\alpha\beta}Z_\beta
\right\},
\end{equation}
\end{linenomath*}
}
where $Z$ represents the \cPZ\ boson and $F_{\mu\alpha}$ represents
the electromagnetic field tensor.
The coefficients $f_i^\cPgg$ and $f_i^\cPZ$ correspond to couplings
$\cPgg^{(\ast)}\cPZ\cPZ$ and $\cPZ^{(\ast)}\cPZ\cPZ$,
respectively. All the operators in Eq.~(\ref{eq:atgc_lagrangian}) are
Lorentz-invariant and $\mathrm{U}(1)_{\mathrm{EM}}$ gauge-invariant,
but not invariant under
$\mathrm{SU}(2)_{\mathrm{L}}\times \mathrm{U}(1)_{\mathrm{Y}}$ gauge
symmetry. The terms corresponding to $f_4^\mathrm{V}$ parameters
violate the CP symmetry, while the terms corresponding to
$f_5^\mathrm{V}$ parameters conserve CP.

To avoid unitarity violation at energies above the scale ($\Lambda$) of
new physics, the Lagrangian of Eq.~(\ref{eq:atgc_lagrangian}) can
be modified with form factors of the type
$1/\left(1+\hat{s}/\Lambda\right)^n$, where $\sqrt{\hat{s}}$ is the
effective center-of-mass energy of the collision. No form-factor
scaling is used in this analysis. This allows to provide results without any bias
that can arise due to a particular choice of the
form-factor energy dependence.

Previous studies of neutral anomalous triple gauge couplings were
performed at LEP2~\cite{Alcaraz:2006mx},
Tevatron~\cite{PhysRevLett.100.131801}, and
LHC~\cite{:2012rg,Khachatryan:2014dia,:2012kg}. No deviation from the SM
expectation has been
observed so far, and the best limits were set by the
LHC measurements based on integrated luminosities of about 5\,(19.6)\fbinv at 7\,(8)\TeV.

\subsection{Limits from the \texorpdfstring{$\cPZ\cPZ\to2\Pl2\PGn$}{ZZ to 2 l 2 nu} channel}

In the following, we extract limits on the neutral triple gauge
couplings $\mathrm{V}^{(\ast)}\cPZ\cPZ$ with the same datasets at 7
and 8\TeV as used for the \cPZ\cPZ\ cross section measurement
described in the previous section. Limits on the four $f_i^\mathrm{V}$
parameters are set by comparing the data with theoretical
predictions.

\begin{figure}[htbp]
  \centering
  \includegraphics[width=0.48\textwidth]{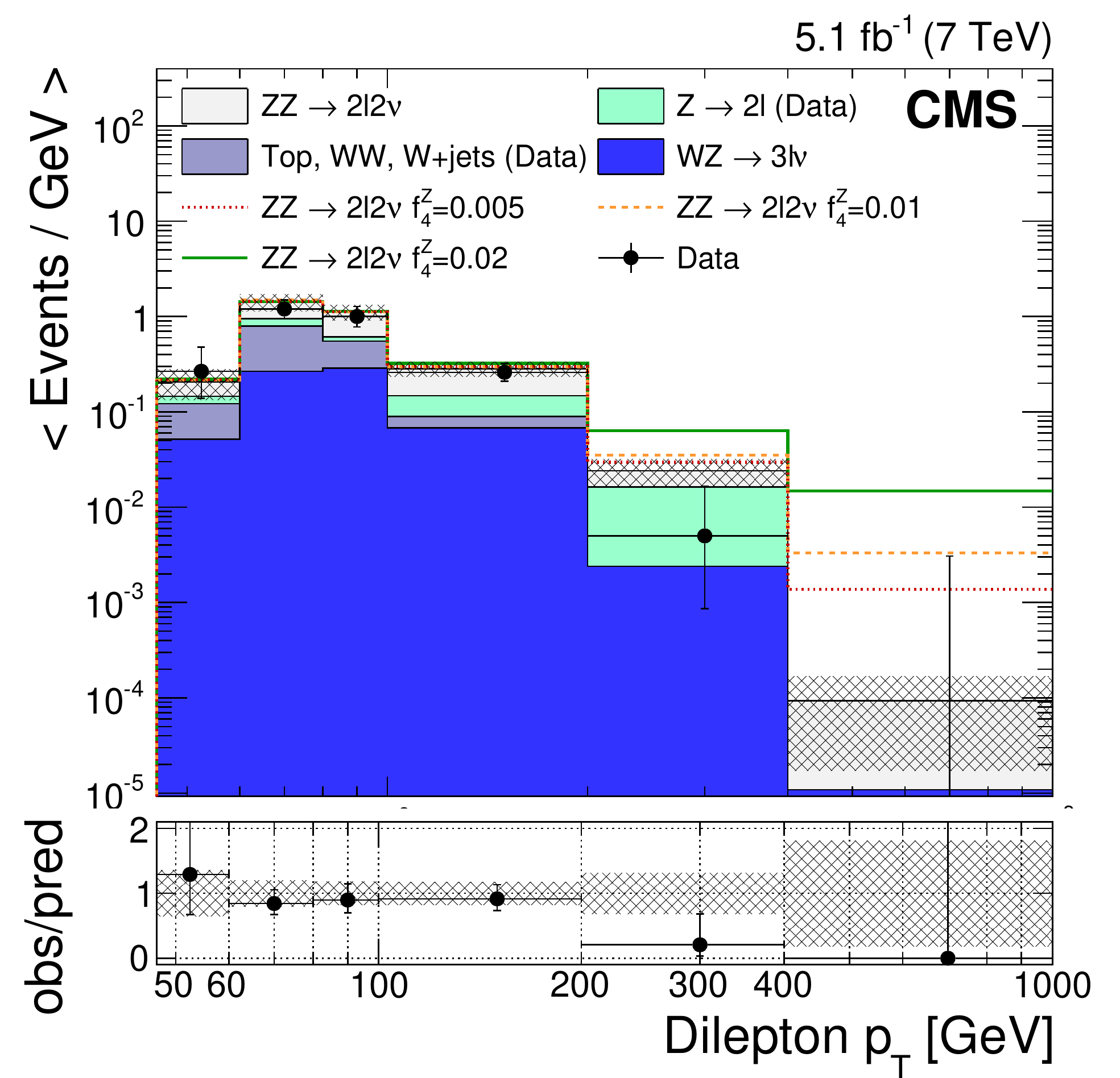}
  \includegraphics[width=0.48\textwidth]{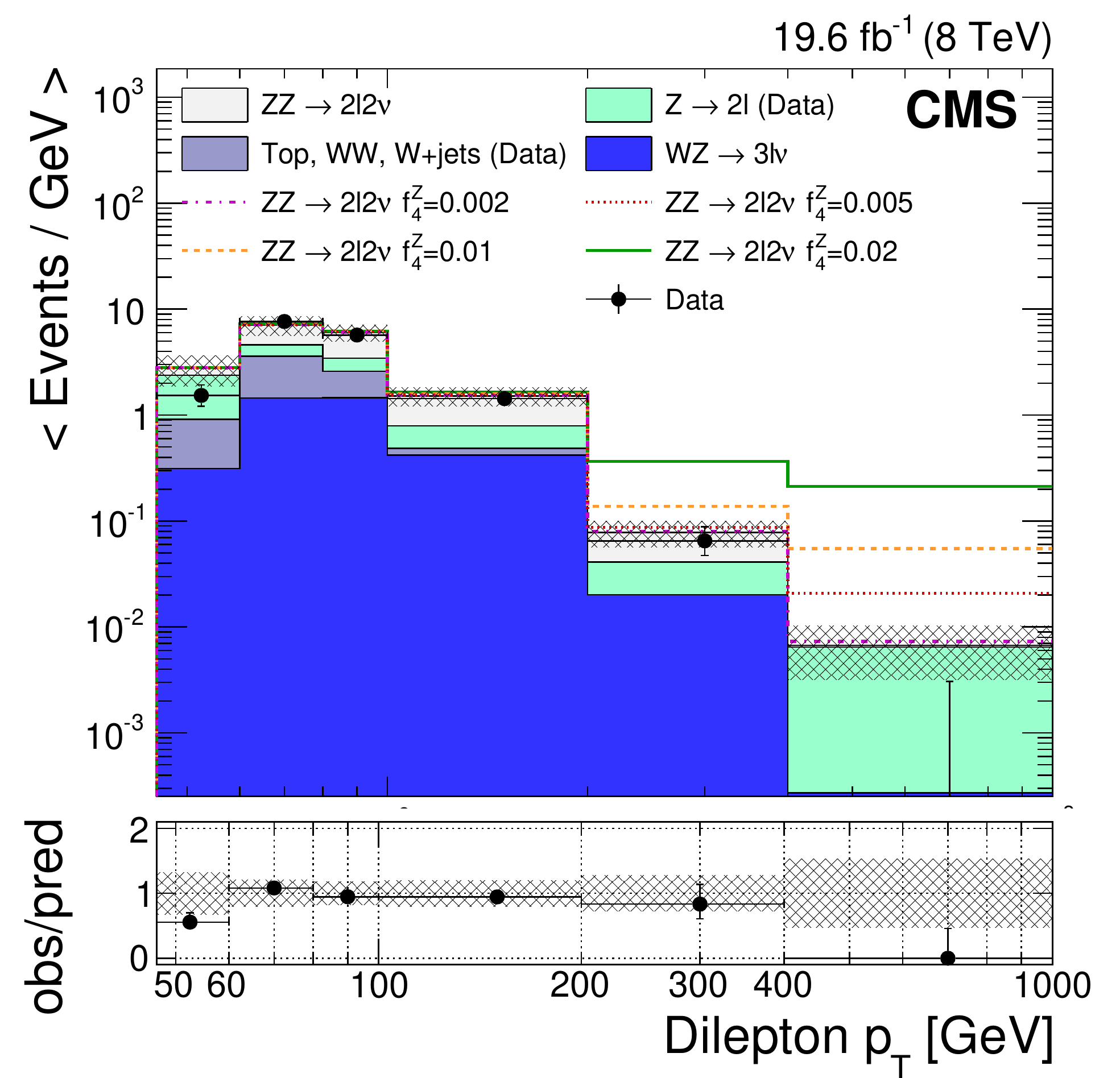}
  \caption{Dilepton ($\Pl = \Pe, \Pgm$) transverse momentum
    distributions at 7\TeV (\cmsLeft) and 8\TeV (\cmsRight).
    The DY and \PW\PW, \PW+jets, and top backgrounds are estimated
    from control samples in data. The gray error band includes
    statistical and systematic uncertainties in the predicted
    yields. In the bottom plots, vertical error bars and bands are
    relative to the total predicted yields. In all plots, horizontal
    error bars indicate the bin width.}
  \label{fig:ll_zpt_rebin_final_onlyF4Z}
\end{figure}

Figure~\ref{fig:ll_zpt_rebin_final_onlyF4Z} shows the charged dilepton
\pt distribution after the full selection described in
Table~\ref{tab:selectioncuts}, in data and simulation, including
\SHERPA samples with different values of the $f_4^\cPZ$ parameter.
The contribution from the anomalous couplings enhances the
high-\pt region of the distribution. The charged dilepton \pt is thus a good
observable to probe for the presence of ATGCs.
The DY and nonresonant backgrounds are estimated from data as
described above.
The SM \cPZ\cPZ\ process is simulated here using the
\MADGRAPH sample described in Section~\ref{sec:CMS},
with NLO QCD corrections computed with {\MCFM} and NLO EW corrections
from Ref.~\cite{Bierweiler:2013dja}.
The contribution of the ATGCs is obtained from the \SHERPA samples
mentioned above, by subtracting the SM \SHERPA contribution to the charged
dilepton \pt, and is summed to the \MADGRAPH \cPZ\cPZ\ distribution.
The interference of the ATGC signal and the SM \cPZ\cPZ\ production is
included, except for $\pt(\cPZ)<200\GeV$, which has a
negligible impact on the limits.
The expected signal yields in each \pt bin are interpolated between
different values of the ATGC coupling parameters using a second-degree
polynomial, since the signal cross section depends quadratically
on such parameters.

The limits are calculated with a profile likelihood
method. We set one-dimensional limits on the four parameters, \ie varying
independently a
single parameter at a time, while fixing the other three to zero.
The 95\% CL one-dimensional limits on the four parameters
are reported in Table~\ref{tab:limits_all} for 7\TeV, 8\TeV, and
combined datasets.
The observed exclusion limits are about one standard deviation tighter
than the expected ones, which is attributed primarily to the observed
deficit of events in the highest bin of dilepton \pt.
The limits set are of comparable sensitivity to those previously
obtained by CMS in the
$4\Pl$ channel~\cite{:2012rg,Khachatryan:2014dia}.

\begin{table*}[htb]
\centering
\topcaption{Summary of 95\% CL intervals for the neutral ATGC
  coefficients, set by the $2\Pl2\PGn$ final states using the 7 and 8\TeV CMS
  datasets. The expected 95\% CL intervals obtained using the 7 and
  8\TeV simulated samples are also shown. No form factor is used.}
\label{tab:limits_all}
 \renewcommand{\arraystretch}{1.2}
\cmsTable{
\begin{tabular}{c|cccc}
  \hline
  \multirow{2}{*}{Dataset} & \multirow{2}{*}{$f_4^\cPZ$}        & \multirow{2}{*}{$f_4^{\cPgg}$}
                           & \multirow{2}{*}{$f_5^\cPZ$}        & \multirow{2}{*}{$f_5^{\cPgg}$}        \\
                      &                   &                   &                   &                  \\
  \hline
  7\TeV    & [-0.010; 0.011]   & [-0.012; 0.013]   & [-0.010; 0.010]   & [-0.013; 0.013]  \\
  \hline
  8\TeV    & [-0.0033; 0.0037] & [-0.0044; 0.0038] & [-0.0033; 0.0035] & [-0.0039; 0.0043]\\
  \hline
  Combined & [-0.0028; 0.0032] & [-0.0037; 0.0033] & [-0.0029; 0.0031] & [-0.0033; 0.0037]\\
  \hline
  Expected (7 and 8\TeV)
           & [-0.0048; 0.0051] & [-0.0060; 0.0053] & [-0.0048; 0.0050] & [-0.0057; 0.0062]\\
  \hline
\end{tabular}
}
\end{table*}

\subsection{Combined limits from the \texorpdfstring{$\cPZ\cPZ\to4\Pl$}{ZZ to 4 lepton} and \texorpdfstring{$\to2\Pl2\PGn$}{2 l 2 nu} channels}

We proceed with the combination of the results of the
previously published $\cPZ\cPZ \to 4\Pl$
analyses~\cite{:2012rg,Khachatryan:2014dia} with the present
results. In doing this, the published analysis of the
$4\Pl$ ($\Pl=\Pe,\Pgm$) channel is unchanged,
except that NLO EW corrections to the SM $\cPZ\cPZ \to
4\Pl$ background are included in the same way as in the present
analysis. We use a profile likelihood method to calculate the 95\%
CL one-dimensional intervals for the four parameters, combining the
data in the $4\Pl$ and $2\Pl2\PGn$ channels, at 7 and 8\TeV. The
systematic uncertainties in the signal and diboson background cross
sections, in the integrated luminosity, and in the lepton efficiencies
are treated as fully correlated between the two channels.
Table~\ref{tab:limits_4l_2l2n} shows the intervals obtained
by combining the four separate data sets.
The combined analysis
improves the sensitivity of the two separate channels, and the limits
are more stringent than all the results published to date.

\begin{table*}[htb]
\centering
\topcaption{Summary of 95\% CL intervals for the neutral ATGC
  coefficients, set by the combined analysis of $4\Pl$ and $2\Pl2\PGn$
  final states. The intervals obtained separately by the two analyses using
  the 7 and 8\TeV CMS data sets are shown, as well as their
  combination. The expected 95\% CL intervals obtained using the 7
  and 8\TeV simulated samples of both analyses are also shown. No form
  factor is used.}
\label{tab:limits_4l_2l2n}
\cmsTable{
\begin{tabular}{c|cccc}
  \hline
  \multirow{2}{*}{Dataset} & \multirow{2}{*}{$f_4^\cPZ$}        & \multirow{2}{*}{$f_4^\cPgg$}
                           & \multirow{2}{*}{$f_5^\cPZ$}        & \multirow{2}{*}{$f_5^\cPgg$}        \\
                      &                   &                   &                   &                  \\
  \hline
  7\TeV, $4\Pl$      & [-0.010; 0.011]   & [-0.012; 0.013]   & [-0.011; 0.011]   & [-0.013; 0.013]  \\
  7\TeV, $2\Pl2\PGn$ & [-0.010; 0.011]   & [-0.012; 0.013]   & [-0.010; 0.010]   & [-0.013; 0.013]  \\
  8\TeV, $4\Pl$      & [-0.0041; 0.0044] & [-0.0052; 0.0048] & [-0.0041; 0.0040] & [-0.0048; 0.0045]\\
  8\TeV, $2\Pl2\PGn$ & [-0.0033; 0.0037] & [-0.0044; 0.0038] & [-0.0033; 0.0035] & [-0.0039; 0.0043]\\
  \hline
  Combined            & [-0.0022; 0.0026] & [-0.0029; 0.0026] & [-0.0023; 0.0023] & [-0.0026; 0.0027]\\
  \hline
  Expected            & \multirow{2}{*}{[-0.0036; 0.0039]} & \multirow{2}{*}{[-0.0046; 0.0041]}
                      & \multirow{2}{*}{[-0.0036; 0.0037]} & \multirow{2}{*}{[-0.0043; 0.0043]}      \\
  ($4\Pl$ and $2\Pl2\PGn$, 7 and 8\TeV)
                      &                   &                   &                   &                  \\
  \hline
\end{tabular}
}
\end{table*}

\section{Summary}
\label{sec:conclusions}

We have measured the \cPZ\cPZ\ production cross section in the $2\Pl2\PGn$
channel in proton-proton collisions
at center-of-mass energies of 7 and 8\TeV.
The data samples selected for the study correspond to an integrated
luminosity of
5.1\,(19.6)\fbinv
at 7\,(8)\TeV.
We have measured
\begin{linenomath*}
$$\sigma(\Pp\Pp \to \cPZ\cPZ) = 5.1_{-1.4}^{+1.5}\stat\,_{-1.1}^{+1.4}\syst\pm 0.1\lum\unit{pb}$$
\end{linenomath*}
at 7\TeV, and
\begin{linenomath*}
$$\sigma(\Pp\Pp \to \cPZ\cPZ) = 7.2_{-0.8}^{+0.8}\stat\,_{-1.5}^{+1.9}\syst\pm 0.2\lum\unit{pb}$$
\end{linenomath*}
at 8\TeV,
in agreement with theory calculations,
$6.2^{+0.3}_{-0.2}$\unit{pb} ($7.6^{+0.4}_{-0.3}$\unit{pb}) at 7\,(8)\TeV,
which include NLO QCD corrections~\cite{bib:MCFM} and NLO EW
corrections~\cite{Bierweiler:2013dja,Baglio:2013toa}.
The selected data have also been
analyzed to search for ATGCs involving the \cPZ\cPZ\ final state.
In the absence of any observation of new physics,
we have set the most stringent limits to date on the relevant
ATGC parameters.
In addition, by combining the selected data with the CMS data for the
four-charged-lepton final state
we have set even tighter constraints.

\begin{acknowledgments}
\hyphenation{Bundes-ministerium Forschungs-gemeinschaft Forschungs-zentren}

We wish to thank our theoretician colleague Tobias Kasprzik for providing the numerical calculations of the next-to-leading-order electroweak corrections to the \cPZ\cPZ\ and \PW\cPZ\ processes.

\hyphenation{Bundes-ministerium Forschungs-gemeinschaft Forschungs-zentren} We congratulate our colleagues in the CERN accelerator departments for the excellent performance of the LHC and thank the technical and administrative staffs at CERN and at other CMS institutes for their contributions to the success of the CMS effort. In addition, we gratefully acknowledge the computing centers and personnel of the Worldwide LHC Computing Grid for delivering so effectively the computing infrastructure essential to our analyses. Finally, we acknowledge the enduring support for the construction and operation of the LHC and the CMS detector provided by the following funding agencies: the Austrian Federal Ministry of Science, Research and Economy and the Austrian Science Fund; the Belgian Fonds de la Recherche Scientifique, and Fonds voor Wetenschappelijk Onderzoek; the Brazilian Funding Agencies (CNPq, CAPES, FAPERJ, and FAPESP); the Bulgarian Ministry of Education and Science; CERN; the Chinese Academy of Sciences, Ministry of Science and Technology, and National Natural Science Foundation of China; the Colombian Funding Agency (COLCIENCIAS); the Croatian Ministry of Science, Education and Sport, and the Croatian Science Foundation; the Research Promotion Foundation, Cyprus; the Ministry of Education and Research, Estonian Research Council via IUT23-4 and IUT23-6 and European Regional Development Fund, Estonia; the Academy of Finland, Finnish Ministry of Education and Culture, and Helsinki Institute of Physics; the Institut National de Physique Nucl\'eaire et de Physique des Particules~/~CNRS, and Commissariat \`a l'\'Energie Atomique et aux \'Energies Alternatives~/~CEA, France; the Bundesministerium f\"ur Bildung und Forschung, Deutsche Forschungsgemeinschaft, and Helmholtz-Gemeinschaft Deutscher Forschungszentren, Germany; the General Secretariat for Research and Technology, Greece; the National Scientific Research Foundation, and National Innovation Office, Hungary; the Department of Atomic Energy and the Department of Science and Technology, India; the Institute for Studies in Theoretical Physics and Mathematics, Iran; the Science Foundation, Ireland; the Istituto Nazionale di Fisica Nucleare, Italy; the Ministry of Science, ICT and Future Planning, and National Research Foundation (NRF), Republic of Korea; the Lithuanian Academy of Sciences; the Ministry of Education, and University of Malaya (Malaysia); the Mexican Funding Agencies (CINVESTAV, CONACYT, SEP, and UASLP-FAI); the Ministry of Business, Innovation and Employment, New Zealand; the Pakistan Atomic Energy Commission; the Ministry of Science and Higher Education and the National Science Centre, Poland; the Funda\c{c}\~ao para a Ci\^encia e a Tecnologia, Portugal; JINR, Dubna; the Ministry of Education and Science of the Russian Federation, the Federal Agency of Atomic Energy of the Russian Federation, Russian Academy of Sciences, and the Russian Foundation for Basic Research; the Ministry of Education, Science and Technological Development of Serbia; the Secretar\'{\i}a de Estado de Investigaci\'on, Desarrollo e Innovaci\'on and Programa Consolider-Ingenio 2010, Spain; the Swiss Funding Agencies (ETH Board, ETH Zurich, PSI, SNF, UniZH, Canton Zurich, and SER); the Ministry of Science and Technology, Taipei; the Thailand Center of Excellence in Physics, the Institute for the Promotion of Teaching Science and Technology of Thailand, Special Task Force for Activating Research and the National Science and Technology Development Agency of Thailand; the Scientific and Technical Research Council of Turkey, and Turkish Atomic Energy Authority; the National Academy of Sciences of Ukraine, and State Fund for Fundamental Researches, Ukraine; the Science and Technology Facilities Council, UK; the US Department of Energy, and the US National Science Foundation.

Individuals have received support from the Marie-Curie program and the European Research Council and EPLANET (European Union); the Leventis Foundation; the A. P. Sloan Foundation; the Alexander von Humboldt Foundation; the Belgian Federal Science Policy Office; the Fonds pour la Formation \`a la Recherche dans l'Industrie et dans l'Agriculture (FRIA-Belgium); the Agentschap voor Innovatie door Wetenschap en Technologie (IWT-Belgium); the Ministry of Education, Youth and Sports (MEYS) of the Czech Republic; the Council of Science and Industrial Research, India; the HOMING PLUS program of Foundation for Polish Science, cofinanced from European Union, Regional Development Fund; the Compagnia di San Paolo (Torino); the Consorzio per la Fisica (Trieste); MIUR project 20108T4XTM (Italy); the Thalis and Aristeia programs cofinanced by EU-ESF and the Greek NSRF; and the National Priorities Research Program by Qatar National Research Fund.
\end{acknowledgments}

\bibliography{auto_generated}

\cleardoublepage \appendix\section{The CMS Collaboration \label{app:collab}}\begin{sloppypar}\hyphenpenalty=5000\widowpenalty=500\clubpenalty=5000\textbf{Yerevan Physics Institute,  Yerevan,  Armenia}\\*[0pt]
V.~Khachatryan, A.M.~Sirunyan, A.~Tumasyan
\vskip\cmsinstskip
\textbf{Institut f\"{u}r Hochenergiephysik der OeAW,  Wien,  Austria}\\*[0pt]
W.~Adam, T.~Bergauer, M.~Dragicevic, J.~Er\"{o}, M.~Friedl, R.~Fr\"{u}hwirth\cmsAuthorMark{1}, V.M.~Ghete, C.~Hartl, N.~H\"{o}rmann, J.~Hrubec, M.~Jeitler\cmsAuthorMark{1}, W.~Kiesenhofer, V.~Kn\"{u}nz, M.~Krammer\cmsAuthorMark{1}, I.~Kr\"{a}tschmer, D.~Liko, I.~Mikulec, D.~Rabady\cmsAuthorMark{2}, B.~Rahbaran, H.~Rohringer, R.~Sch\"{o}fbeck, J.~Strauss, W.~Treberer-Treberspurg, W.~Waltenberger, C.-E.~Wulz\cmsAuthorMark{1}
% ...
\vskip\cmsinstskip
% ...
1:~~Also at Vienna University of Technology, Vienna, Austria\\
2:~~Also at CERN, European Organization for Nuclear Research, Geneva, Switzerland\\
% ...

\end{sloppypar}
\end{document}